\documentstyle[preprint,prb,eqsecnum,aps]{revtex}

\begin{document}
\draft
\preprint{HEP/123-qed}
\title{Density Matrix Renormalization Group study of the polaron 
problem in the Holstein model}
\author{Eric Jeckelmann and Steven R. White}
\address{Department of Physics and Astronomy, University of California,
Irvine, CA 92697.
}
\date{\today}
\maketitle
\begin{abstract}
We propose a new density matrix renormalization group
(DMRG) approach to study lattices including bosons.
The key to the new approach is an exact mapping
of a boson site containing $2^N$ states to $N$
pseudo-sites, each with 2 states.
The pseudo-sites can be viewed as the binary digits of
a boson level. 
We apply the pseudo-site DMRG method to the 
polaron problem in the one- and two-dimensional Holstein 
models.
Ground state results are presented for a wide range of 
electron-phonon coupling strengths and phonon frequencies
on lattices large enough (up to 80 sites in one dimension and 
up to 20$\times$20 sites in two dimensions)   
to eliminate finite size effects, 
with up to 128 phonon states per phonon mode. 
We find a smooth but quite abrupt crossover from a
quasi-free electron ground state with a slightly
renormalized mass at weak electron-phonon 
coupling to a polaronic ground state with a large 
effective mass at strong coupling,
in agreement with previous studies.

\end{abstract}

\narrowtext

\section{Introduction}

The density matrix renormalization group (DMRG) method 
\cite{whi92,whi96}
has proved to be a very successful numerical technique for
studying spin and fermion lattice models with short-range
interactions in low dimensions.
Although the DMRG algorithm can easily be generalized to treat 
systems including bosons, calculations are often not practical.
As for exact diagonalizations, this is due to the difficulty in
dealing with the large (in principle, infinite) dimension of the
Hilbert space for bosons.
Although the problem is less severe in DMRG than in exact 
diagonalizations, 
applications of DMRG to boson systems have been
restricted to problems for which one needs to consider 
at most about a dozen states
for each boson \cite{pai96,car96,car97}.

In this paper, we present a new approach for dealing
with large bosonic Hilbert spaces with DMRG.
The basic idea is to transform each boson
site into several artificial interacting 2-state sites 
(pseudo-sites) and then to use DMRG techniques to treat 
this interacting system. 
DMRG is much better able to handle several 2-state sites
rather than one many-state site.
Although this procedure 
introduces some complications in a DMRG program, 
the pseudo-site approach is more efficient and
allows us to keep many more states in each bosonic Hilbert space
than the approach used in earlier
works \cite{pai96,car96,car97}. 

To test our method, we have studied the polaron problem, 
the self-trapping of an electron by a localized lattice 
deformation, 
in the Holstein model \cite{hol59} in one and two dimensions. 
We consider a single electron
on a lattice with oscillators of frequency $\omega$ at each
site representing dispersionless optical phonon modes
and a coupling between the electron density and oscillator
displacements $q_\ell = b^\dag_\ell + b_\ell $, 
where $b^\dag_\ell$ and $b_\ell$ are the usual
boson creation and annihilation operators.
The Hamiltonian is given by 
\begin{equation}
H =  
\omega \, \sum_\ell b^\dag_\ell b_\ell 
- g \, \omega \, \sum_\ell \left (b^\dag_\ell + b_\ell \right ) 
n_\ell
- t \sum_{\langle\ell,m\rangle} \left (c^\dag_{m}
c_{\ell} + c^\dag_{\ell} c_{m} \right )  \, ,
\label{eq:ham}
\end{equation}
where $c^\dag_\ell$ and $c_\ell$ are electron creation and annihilation 
operators, $n_\ell = c^\dag_\ell c_\ell$ and $t$ is the hopping integral.
$g$ is a dimensionless electron-phonon coupling constant.
(When comparing results readers should be aware 
that notations for model parameters, especially $g$,
differ in other papers.)
A summation over $\ell$ or $\langle\ell,m\rangle$
means a sum over all sites or over
all bonds between nearest-neighbor sites 
in a chain of length $L$ or a square lattice of size $L\times L$. 
Only open systems have been considered because the 
DMRG method usually performs much
better in this case than for periodic boundary conditions.

The polaron problem has been extensively studied using 
variational methods \cite{var}, 
quantum Monte Carlo simulations \cite{rae83,kor97}, 
exact diagonalizations \cite{ran92,ale94,mar95,wel96,cap96} 
and perturbation theory \cite{mar95,cap96,ste96}. 
It is known that a rather sharp crossover occurs between
a quasi-free-electron ground state with a slightly 
renormalized mass at weak electron-phonon coupling, and a
polaronic ground state with a narrow band-width at strong coupling.
However, despite these considerable theoretical efforts,
the physics of this self-trapping transition 
is not fully understood.
Previous studies have been limited either to small systems
or to a particular regime of parameters $g$ and $\omega/t$ 
or by a severely truncated phononic 
Hilbert space or by uncontrolled approximations.
With the DMRG method, we have been able to study
the one-electron ground state of the Holstein model for all 
regimes of parameters $\omega/t$ and $g$ 
on large lattices and with great accuracy. 
In this work we report and discuss some ground state results 
which shows the self-trapping crossover, such as
electron-lattice displacement correlation functions,
electronic kinetic energy, and effective mass.

This paper is organized as follows: 
in the next section, we present our new pseudo-site method for 
bosons. 
In Section 3 we describe how we apply this method to the 
Holstein model. 
Most results for the polaron problem are presented and discussed in 
Section 4. 
In Section~5 we explain how we have computed the effective mass
of electrons and polarons and present these results.
Finally, Section~6 contains our conclusions.

\section{DMRG for boson systems}

In the DMRG method, the lattice is broken up into blocks made
of one or several sites 
and Hilbert spaces representing blocks are truncated 
(for more details, see Refs. 1 and 2).
In each block one keeps  only the $m$ most important
states for forming the ground state 
(or low-energy eigenstates) of the full system.
A step of the DMRG algorithm is the process
of forming a new block by adding a site to a block 
obtained in a previous step. 
To find the $m$ optimal states of the new enlarged block,
one has to find the ground state of an 
effective Hamiltonian in a superblock made of two blocks and two
sites and then 
to diagonalize a density matrix on the new block.
If $n$ is the number of states on a site, 
the computer memory storage needed to perform these
tasks increases as $n^2 \, m^2$,  
while the number of operations  goes roughly as $n^3 \, m^3$.  

The difficulty in applying the DMRG to boson systems is the large
number of states on a site.
In principle, this
number is infinite and for numerical calculations one has to
truncate this space and keep a finite number $M$ of states per
boson.  
In a standard implementation of the DMRG method for boson
systems, each boson forms one lattice site ($n \sim M$)
and thus memory and CPU time requirements increase 
as $M^2$ and $M^3$, respectively.
For many interesting problems, such as the Holstein polaron discussed in
this paper, one needs to keep a large number of states per boson
sites ($M \approx 10-100$) to reduce errors due to
the truncation of bosonic Hilbert spaces. 
Therefore, performing such calculations 
requires much more computer resources than DMRG computations for 
otherwise similar 
Heisenberg or Hubbard systems, for which $n = 2-4$.

To understand the basis of our new approach,
it is important to note that, in principle, 
the computer resources used by the DMRG method  
increase linearly with the number of lattice sites
(everything else being equal).
Therefore, 
DMRG performances should be better when individual
lattice sites are defined so that the number of states $n$
is as small as possible (i.e., $n = 2$) even if this
implies an increase in the number of sites in the lattice. 
For instance, in the Hubbard model for fermions, 
we can either use the same site for both spin up and spin down 
fermions or use different sites for fermions of different spins. 
In the first case, the Hilbert
space contains $n = 4$ states 
($|0\rangle, |\uparrow\rangle, |\downarrow\rangle,
|\uparrow \downarrow \rangle$) per site. 
In the second case, the lattice contains twice
as many sites but the Hilbert space of each site
contains only 2 states 
(($|0\rangle, |\sigma\rangle$, with $\sigma = \uparrow$ 
or $\downarrow$). 
In practice, the  second approach is faster
by a factor of~2.
Also, in a boson-fermion model as the Holstein 
Hamiltonian (\ref{eq:ham}),
a site can have both fermion and boson degrees of freedom,
or one can separate
the boson and fermions into two sites.
We have found that the latter
method is significantly more efficient than the former.
However, it should be kept in mind that DMRG performances
depends  essentially on the number $m$ of states that
one needs to keep per block 
to obtain a desired accuracy, 
the number of iterations needed by the DMRG algorithm to converge
and the possible use of system symmetries. 
All these parameters tend to be unfavorably altered by the partition 
of sites in smaller units and a large increase of $m$ or
of the number of iterations could offset any gain due to 
the reduction of the Hilbert space dimension. 
Nevertheless, experience indicates that it is usually possible
to improve DMRG performances by substituting several sites
with a small Hilbert space for a site with a large Hilbert space.

Therefore, we have developed a method to exactly transform
a boson site into several smaller pseudo-sites. 
Our approach is motivated by a familiar concept:
the representation of a number in binary form.
In this case the number is the boson state index starting at 0.
Each binary digit is represented by a pseudo-site, which can be
occupied (1) or empty (0). 
One can think of these
pseudo-sites as being fermions, but is is simpler to implement
them as hard-core bosons, thus avoiding fermion anticommutation
minus signs.
Thus, for a boson site with $M=2^N$ levels, 
the level with index 0 is represented by $N$ empty
pseudo-sites, while the highest level, $2^N-1$,
is represented by $N$ hard-core bosons on the $N$ pseudo-sites.

To implement this idea,
we first choose a
truncated occupation-number basis \{$|\alpha \rangle, \, 
\alpha~=~0,1,2,...,2^N-1$\}, where 
$b^\dag b \, |\alpha \rangle = \alpha |\alpha \rangle$,
as the finite Hilbert space of a boson site. 
Then, we introduce N pseudo-sites $j=1,...,N$ 
with a 2-dimensional Hilbert 
space $\{|r_j \rangle, r_j = 0,1 \}$ and the operators 
$a^\dag_j, a_j$  such that 
$ a_j |1\rangle = |0 \rangle, \ a_j |0\rangle = 0$
and $a^\dag_j$ is the hermitian conjugate of $a_j$.
These pseudo-site operators have the same properties 
as hard-core boson operators: $a_j a^\dag_j + a^\dag_j a_j = 1$
and operators on different pseudo-sites commute.
The one-to-one mapping
between a boson level $|\alpha \rangle$ and the $N$-pseudo-site
state $|r_1, r_2, ..., r_N \rangle$
is given by the relation
\begin{equation}
\alpha = \sum_{j=1}^{N}  \, 2^{j-1} \, r_j \,.
\label{eq:map}
\end{equation}

The next step is to write all boson operators in terms  of
pseudo-site operators. 
It is obvious that the boson number operator is 
given by
\begin{equation}
N_b \, = \, b^\dag b \, = \, \sum_{j=1}^{N}  
\, 2^{j-1} \, a^\dag_j a_j.
\end{equation}
Unfortunately, other boson operators take a more complicated form 
in the pseudo-site representation.
Typically, they are represented by a sum over $ \sim M$ terms.
They can easily be  determined from the
definition of the mapping (\ref{eq:map}) and the properties of 
boson and hard-core boson operators. 
As an example, we show here how to calculate the representation of $b^\dag$.
First, we write $b^\dag = B^\dag \,\, \surd \overline{N_b+1}$,
where $B^\dag |\alpha\rangle = |\alpha+1\rangle$.
The pseudo-site operator representation of the second term is
\begin{equation}
\surd \overline{N_b+1} \,  = \, 
\sum_{\alpha=0}^{M-1} \surd \overline{\alpha+1} 
\, \, P_1(r_1) \, P_2(r_2) \, ... \, P_N(r_N)  \, ,
\end{equation}
where $P_j(1) = a^\dag_j a_j$, $P_j(0) = a_j a^\dag_j$ and
the $r_j$ ($j=1,..,N$) are given by the mapping~(\ref{eq:map}).
For $B^\dag$ we find
\begin{equation}
B^\dag  \,  = \, 
a^\dag_1 + a^\dag_2 \, a_1 + a^\dag_3 \, a_2 \, a_1 + ...
+ a^\dag_N \, a_{N-1} \, a_{N-2} \, ... \,a_1.
\end{equation}
The representation of $b^\dag$ for any number of pseudo-sites
$N$ is given by the product of these two operators.  
For instance, for $N=2$ pseudo-sites 
\begin{equation}
b^\dag \, = \, a^\dag_1 +  \sqrt{2} \, \, a^\dag_2 \, a_1 + \left (
\sqrt{3}-1 \right ) a^\dag_1 \, a^\dag_2 \,  a_2.
\label{eq:bdag}
\end{equation}
Other operators can be obtained in a similar way.

We can now substitute $N=\log_2(M)$ pseudo-sites for each boson site
in the lattice and rewrite the system Hamiltonian and other operators
in terms of the pseudo-site operators.
Once this transformation has been done, 
DMRG algorithms can be used to calculate 
the system properties.
For instance, if one would like to find the ground state of
an oscillator in a linear potential
$H \, = \omega b^\dag b - \gamma (b^\dag + b)$
keeping $M$=4 states, 
we would transform this system into a 2-site hard-core boson system
with the Hamiltonian
\begin{eqnarray}
H \, &=& \, \omega \sum_{j=1}^{2}  
\, 2^{j-1} \, a^\dag_j a_j  \nonumber \\
& & - \gamma \, ( 
a^\dag_1 +  \sqrt{2} \, \, a^\dag_2 \, a_1 + \left (
\sqrt{3}-1 \right ) a^\dag_1 \, a^\dag_2 \,  a_2 \nonumber \\
& & + a_1 +  \sqrt{2} \, \, a^\dag_1 \, a_2 + \left (
\sqrt{3}-1 \right ) a^\dag_2 \,  a_2 \, a_1 )  
\end{eqnarray}
One can easily check that both Hamiltonians 
share the same matrix representation in the basis
\{$|\,\alpha \rangle, \alpha=0,1,2,3$\} and 
\{$|\,r_1, r_2\rangle, r_1=0,1, r_2=0,1$\}, respectively.

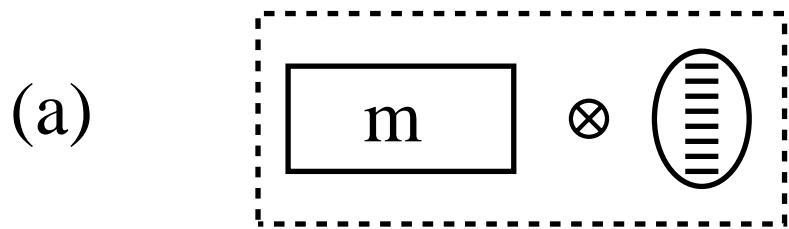
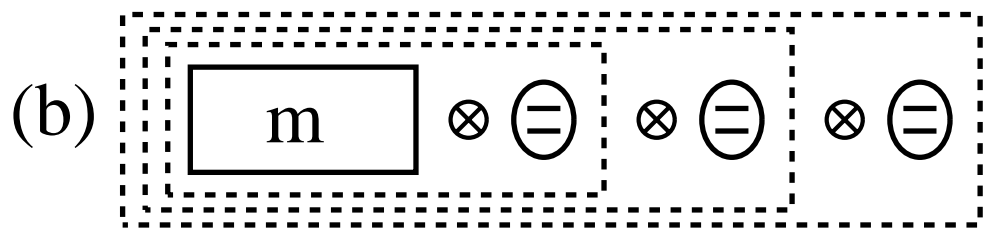
\begin{figure}
\caption{Symbolic representation of a DMRG step for $N$=3 ($M$=8).
In the standard DMRG approach (a), a new block (dashed rectangle) 
is formed by adding a boson site (oval) with its $M=8$ states to the 
initial block (solid rectangle).   
In the pseudo-site approach (b), a new block is made of the previous block 
and one pseudo-site with 2 states. 
It takes $N=3$ steps to make the final block (largest dashed rectangle)
including the initial block and all pseudo-sites,
which is equivalent to the new block in (a).} 
\label{fig:blocks}
\end{figure}

Fig.~\ref{fig:blocks} illustrates the differences between
standard and pseudo-site DMRG approaches.
In the standard approach (Fig.~\ref{fig:blocks}(a)),
a new block is built up 
by adding a boson site
with $M$ states to another block with $m$ states.
Initially, the Hilbert space of the new block
contains $m M$ states and  
is truncated to $m$ states according to the DMRG
method.  
In the pseudo-site approach (Fig.~\ref{fig:blocks}(b)),
we build up a new block by adding one pseudo-site with 2 states
to another block with $m$ states.
The Hilbert space of this new block
contains only $2 m$ states and is also
truncated to $m$ states according to the DMRG
method.  
We have to repeat this step $N$ times until the $M$-state boson
Hilbert space has been added to the original block.
However, at each step we have to manipulate
only a fraction $2/M$ of the bosonic Hilbert space.
It should be noted that the transformation into
pseudo-sites is an exact mapping of truncated
bosonic Hilbert spaces. 
Therefore, the final blocks of both approaches in
Fig.~\ref{fig:blocks} would be equivalent if we did not
truncate the block Hilbert spaces to $m$ states
at each intermediate step. 
Actually, we have never found any significant differences 
between pseudo-site and standard DMRG results but 
it is possible that such differences appear
when the DMRG truncation error
(the error due to the truncation of block Hilbert 
spaces to $m$ states) 
is large enough.

Using the transformation into pseudo-sites
we have implemented and tested several DMRG 
algorithms~\cite{whi92,whi96}.
Generally, implementing a DMRG algorithm for
pseudo-sites is more complicated than
a standard DMRG method.
This artificial transformation 
generates a very complicated Hamiltonian
which includes many (typically $M$) long-range interaction terms
between pseudo-sites. 
Therefore, at each DMRG step one must keep track of and transform
many matrices representing different combinations of pseudo-site
operators used to build up the Hamiltonian and other
operators during following steps.
However, one needs the many pseudo-site operators 
only during the intermediate
steps of Fig.~\ref{fig:blocks}(b).
Once the full bosonic Hilbert space has been added
to the block 
(after the final step in Fig.~\ref{fig:blocks}(b)),
one only needs regular boson operators as in a 
standard DMRG method.
Therefore, in an efficient implementation, 
matrix representations of boson operators should be computed 
from the pseudo-site operator matrices,
which can be discarded after that,
whenever it is possible.
The cost of this operation ($\propto M$ matrix additions)
is small compared
to the cost of keeping track of and transforming
the pseudo-site operator matrices 
($\propto M$ matrix multiplications).

\section{Application to the Holstein model}

We have applied the pseudo-site DMRG method to 
the Holstein model in
different situations: 1 electron in one and two dimensions,
2 electrons and half-filled band systems in one dimension, sometimes
with additional interactions as an on-site impurity
potential or a local electron-electron repulsion
(Hubbard term).
Although most of the discussion in this section applies 
to all these different cases,
all quantitative results provided here regard the 
1-electron system with parameters in the range of
$0.1 \leq \omega/t \leq 4$, $g < 5$.

Several tests have shown that the performance
and stability of the DMRG method applied
to the Holstein model depends greatly
on details of the algorithm used. 
Below we describe the best approach
we have found.
We have used the finite system DMRG algorithm~\cite{whi92}
to calculate properties of a system of fixed size.
However, during the warmup sweep 
we have not used an infinite system algorithm. 
Instead, environment blocks are built up using several sites 
without truncation.
With this procedure the accuracy of the results
after the warmup sweep is very poor, 
but this is not a problem because
in the finite system algorithm
the subsequent iterations
(sweeps back and forth across the lattice) 
can usually make up for a poor quality warmup sweep.
The number of states kept per block $m$ is gradually
increased as one performs the iterations, and we keep
track of the ground state wavefunction from step to step
to reduce the total calculation time~\cite{whi96}.
We have found that it is necessary to 
optimize the approximate ground state 
for each intermediate value of $m$.
This optimization requires 
performing  several iterations (up to 6)
for each intermediate value of $m$
even if the energy gain brought by these sweeps seems
negligible compared to the energy gain which could be made
by increasing $m$ immediately.
Otherwise, the DMRG algorithm does not
truncate block Hilbert spaces optimally
and eventually fails to converge.
We think that these additional iterations are needed
to optimize the delocalization energy of the electron or polaron,
which can be a small fraction of the total energy.
The total number of iterations needed
by the DMRG algorithm to converge
varies greatly and in the worst cases can grow up to 30. 
Although the Holstein Hamiltonian~(\ref{eq:ham}) is 
reflection-symmetric, this symmetry has not been used
in our algorithm.
We have found that using the 
reflection-symmetry can hinder and sometimes
prevent the convergence to the ground state.

With the pseudo-site DMRG method, we have been able
to keep enough states per phonon mode
(up to $M=128$) so that the errors from truncation of the
phonon basis are negligible.
To check that $N$ is large enough, we compute the pseudo-site
density $A_j=\langle a^+_j a_j\rangle$, 
where $\langle...\rangle$ means the expectation value in the ground state,
and extrapolate to find $A_{N+1}$. 
$N$ is chosen so that $A_{N+1}$ is comparable to the DMRG truncation 
error.
Usually $N \leq 6$ ($M\leq 64$) was sufficient.

The polaron problem has an important computational advantage 
as a test case: the number of states $m$ which needs to be kept 
per block is relatively small. 
In the non-interacting case ($g= 0$), one can easily show that 
only two eigenstates of the density matrix have a non-zero weight.
For finite coupling $g$, the DMRG truncation error often vanishes
(within the machine precision $ \approx 10^{-16}$) if
we keep a relatively small number of states. 
Although we need to keep more states when $g$ or 
the system size increases,
we have found that a DMRG truncation error 
smaller than $10^{-14}$ can be reached with at most $m=150$ states
in all our calculations.
This feature has also allowed us to obtain accurate results in quite large
two-dimensional systems.

With the DMRG method, 
the error on the ground state energy is generally
proportional to the DMRG truncation error. 
Therefore, we can calculate the ground state energy
and the truncation error for several values of $m$ and
use a linear fit to extrapolate the result without
truncation error~\cite{whi92}. 
This method gives reliable estimations of the
error on the ground state energy. 
In one dimension we have obtained relative errors in the range 
of $10^{-10}$ to $10^{-16}$ 
depending on the system size and parameters $g$, $\omega/t$.
In two dimensions we have contented ourself with larger errors, from
$10^{-6}$ to $10^{-10}$, to save CPU time
but more accurate results can be obtained.

In the polaronic regime,
the density of states near the ground state becomes very large
(see the discussion in Section~\ref{sec:mass}).
Thus, a small energy error does not guarantee
that we have obtained an accurate  ground state wavefunction.
To estimate the precision of measurements $\langle O \rangle$,
where $O$ is any operator other than the Hamiltonian $H$,
we have used exact relations between expectation values, 
such as symmetry conditions or self-consistence equations.
For instance, the self-consistence equation 
\begin{equation}
\langle b^\dag_\ell + b_\ell \rangle = 2g \langle n_\ell\rangle \, ,
\label{eq:selfc}
\end{equation}
which holds for all eigenvalues of the Holstein Hamiltonian~(\ref{eq:ham}),
gives a local condition on both fermion and boson degrees of freedom. 
For the results presented in this paper
we have typically obtained relative errors smaller than  $10^{-4}$ in 
one dimension and smaller than $10^{-2}$ in two dimensions.
Finally, we point out that the pseudo-site DMRG method perfectly
reproduces exact diagonalization results for the ground state and
lowest excited states of small systems 
like the 2-site Holstein model~\cite{ran92}.

For each value of the parameters $g$ and $w/t$
we have studied systems of different sizes
and checked that finite size effects are negligible
or extrapolated results to an infinite system.
The largest system sizes that we have used to study the 
one-dimensional Holstein model are $L=80$ sites
for $N \leq 5$ ($M \leq 32$) and $L=30$ sites for 
$5< N \leq 7$ ($32 < M \leq 128$).
In two dimensions, we have used square lattices with
up to $20\times20$ sites for $N \leq 3$ ($M \leq 8$)
and up to $12\times12$ sites for $4 < N \leq 6$
($16 < M \leq 64$).
In most cases we could easily study much larger lattices
if we needed to.
However, in the polaronic regime,
the largest system size for which we can compute the
ground state accurately is limited by the finite precision 
of the DMRG method.
We will discuss this point further in Section~\ref{sec:mass}.

The relative small number of states needed for the polaron problem
allow us to carry out some calculations with both the 
standard and pseudo-site approaches and to compare their performances
in terms of CPU time and memory storage.
In test calculations with all parameters equal, 
we have found that performances of both approaches are 
similar for small $M$ but the pseudo-site approach becomes better
for $M \geq 8$. 
The differences between these methods increase very rapidly with $M$, 
as expected, and, more surprisingly, with $m$.
For $M = 32$ and $m = 50$, 
the pseudo-site approach requires only 1/8 of the memory used by 
the standard  approach and is faster by two orders of magnitude.
In real applications, however, we expect the performance
difference between both approaches to be smaller
because of the greatest flexibility and simplicity of a
standard approach.
For instance, $M$ can take any integer value in the
standard approach.
Nevertheless,  when computations become challenging
(for $M \geq 16$ and $m \geq 50$), the pseudo-site approach clearly
outperforms the standard approach.

\section{Results}
Using the numerical method presented in the previous sections, 
we have studied the ground state properties of the 
Holstein Hamiltonian~(\ref{eq:ham})
with a single electron in one and two dimensions. 
In particular, we are interested in the evolution of the 
ground state as a function of the adiabaticity $\omega/t$ and 
of the electron-phonon coupling $g$.
For a weak coupling 
a standard perturbation calculation in $g$ shows that the ground state is 
a quasi-free electron dragging a phonon cloud, 
which slightly renormalizes the electron effective mass.
Note that the weak-coupling regime roughly corresponds to
$g < 1$ and $2g^2\omega < W$, where $W=4t$ in one dimension
and $W=8t$ in two dimensions is the bare electronic band-width. 
The standard strong-coupling theory of the Holstein model~\cite{hir83},
which is based on the Lang-Firsov transformation
and treats the electron hopping term as a perturbation, 
predicts a polaronic ground state with a narrow band-width.
The strong-coupling regime corresponds
to $g > 1$ and $2g^2\omega > W$.
In this section, we present several 
results of pseudo-site DMRG calculations
which show the evolution of the ground states
from the weak to the strong electron-phonon coupling regime
and compare them to the predictions of perturbation
calculations and the results of previous numerical studies.

\subsection{Electronic density}
\label{sec:dens}

For periodic boundary conditions, it is known rigorously
that the ground state energy and wavefunction
are analytic function of the electron-phonon coupling 
$g$~\cite{ger91}.
In particular, no phonon-induced localization
transition (breaking of the translation symmetry)
occurs for finite $g$;
in the ground state the electron is always delocalized 
over the lattice.
Correspondingly, for open chains our DMRG results show that
the electronic density $\langle n_\ell \rangle$ always has the shape
\begin{equation}
n(\ell)\, = \, \frac{2}{L+1-2a} \, \sin^2\left(
\frac{\pi \cdot (\ell-a)}{L+1-2a}\right) 
\label{eq:dens}
\end{equation}
for $1+a \leq \ell \leq L-a$ and $n(\ell)=0$ otherwise, 
where $a$ is an integer number.
This density corresponds to a free particle
in a one-dimensional box made of the sites with
indices $\ell=1+a$ to $\ell=L-a$.
Therefore, the electron is delocalized over the whole lattice,
except for some chain edge effects, in qualitative agreement with 
the exact result for periodic boundary conditions. 
For small coupling $g$, we have found that we obtain
the best fit with $a=0$ as for a free electron.
For stronger couplings better fits can be obtained with
larger values of~$a$.
For instance, Fig.~\ref{fig:density} shows a density obtained
with the DMRG method and the function~(\ref{eq:dens})
for $a=$ 0 and 1.
Even when the best fit is obtained with $a > 0$, the density 
$\langle n_\ell \rangle$ close to the 
chain edges is actually finite but very small. 

On two-dimensional square lattices the electron is also
delocalized over the lattice for all values of the parameters
$g$ and $\omega/t$ that we have investigated. 
For instance, in Fig.~\ref{fig:density2}, 
we show the density $\langle n_{x,y} \rangle$
for a lattice in the strong-coupling regime.
In the weak coupling regime, the electronic density distribution 
has the same shape 
\begin{equation}
n(x,y) \, = \, \frac{4}{(L+1)^2} \, 
\sin^2\left( \frac{\pi \cdot x}{L+1}\right)  
\sin^2\left( \frac{\pi \cdot y}{L+1}\right)  
\label{eq:dens2}
\end{equation}
as the density of a free particle in a two-dimensional box.
As in one dimension, for stronger coupling the density becomes larger
in the middle of the lattice and decreases near the edges, but in this
case we can not fit the density $\langle n_{x,y} \rangle$ 
with Eq.~(\ref{eq:dens2}) and a renormalized system size.

\subsection{Electron-lattice correlations}
Some ground state properties can easily be studied in terms of
static correlation functions $\langle n_i \,q_j \rangle$ between
the electron position and the oscillator displacement
$q_j = b^\dag_j + b_j$. 
These correlations indicate the strength
(for $i=j$) of the electron-induced lattice deformation
and its spatial extent. 
In the non-interacting case ($g=0$) they are uniformly zero.
Fig.~\ref{fig:correl1} shows the normalized correlation functions
$\chi_{10,j}= \langle n_{10}\, q_j \rangle/\langle n_{10} \rangle$ 
for several parameters $\omega/t$ and $g$ in 20-site chains.
For parameters close to the weak-coupling regime 
(Fig.~\ref{fig:correl1}(a) and (c))
the amplitude of $\chi_{10,j}$ is smaller than the 
quantum lattice fluctuations, which are given
by the zero-point fluctuations of each phonon mode
$\sigma_q \approx 1$. 
Therefore, these correlations do not show a lattice deformation
which could trap an electron because the sign of
the effective lattice potential seen by the electron
fluctuates.
They are merely the signature of a phonon cloud following the electron.
For parameters close to the strong-coupling regime
(Fig.~\ref{fig:correl1}(b) and (d)), 
the amplitude of $\chi_{10,j}$ is larger than these quantum 
lattice fluctuations. 
In these cases, we really observe a lattice deformation generating 
a local attractive potential which is likely to trap the electron.

We observe similar features in two-dimensional lattices.
Fig.~\ref{fig:correl2} shows a normalized correlation function
$\chi(x,y) = \langle n_{8,8}\, q_{x,y} \rangle/\langle n_{8,8} \rangle$
in the weak-coupling regime.
The amplitude of $\chi(x,y)$ is much smaller than
quantum lattice fluctuations $\sigma_q \approx 1$.
In Fig.~\ref{fig:correl2b} we show a similar correlation function,
$\chi(x,y) = \langle n_{5,5}\, q_{x,y} \rangle/\langle n_{5,5} \rangle$,
in the strong-coupling regime.
In this case the amplitude of the lattice
deformation generated by the electron
is clearly larger than the zero-point lattice fluctuations.

In the weak-coupling limit we observe an exponential decay 
of correlations between electron position and lattice deformation.
We find an good agreement between our DMRG results and weak-coupling 
perturbation results for all phonon frequencies $\omega/t$, 
even in the non-adiabatic regime ($\omega/t > 1$) where 
the correlations decrease very fast.
In the adiabatic ($\omega/t << 1$) weak-coupling limit, 
the lattice deformation extends over many 
sites (Fig.~\ref{fig:correl1}(a)).
When $g$ or $\omega/t$ increases,
the spatial extent of the lattice deformation decreases.
In the strong coupling limit, the ground state becomes
"superlocalized" in the sense that any operator measuring
a correlation between the electron and a phonon vanishes
unless the correlation is measured on the same site~\cite{rae83}.
In particular, one finds $\langle n_i q_j \rangle \sim \delta_{ij}$
(see Fig.~\ref{fig:correl1}(d) and  Fig.~\ref{fig:correl2b}).
The variation of the lattice deformation extent
as a function of $\omega/t$ can easily
be understood as a retardation effect.
For small $\omega/t$, phonons are much slower than the electron
and thus phonon modes which are excited by the passage
of the electron take a long time to relax.
Therefore, we can observe a lattice deformation far away from
the current position of the electron. 
In the anti-adiabatic limit ($\omega/t >>1$), lattice fluctuations
are fast and a lattice deformation relaxes quickly following
the slow electronic motion.
Thus, we can observe
a lattice deformation only in the vicinity of the electron. 

It should be kept in mind that these correlations only 
show expectation values of the lattice displacements $q_\ell$
with respect to an instantaneous electron position.
They do not show the electron density distribution for a 
specific frozen lattice configuration. 
Therefore, these results alone are not evidence for the formation
of a self-trapped electronic state and they give no information
regarding the electron density distribution within a polaron. 
To obtain this information we should compute 
$\langle P_i \,n_j\rangle$,
where $P_i$ projects the phonon states onto a particular 
lattice configuration representing a polaron centered 
on site $i$.
Unfortunately, we do not know the operator $P_i$.

\subsection{Self-trapping crossover}
Previous numerical studies have shown that there is a critical
value of the electron-phonon coupling above which
self-trapping of the electron by a local
lattice distortion does 
occur~\cite{rae83,ran92,wel96,cap96}.
One should keep in mind that no localization of the ground state 
wavefunction is involved in self-trapping. 
Therefore, a smooth crossover from a quasi-free electron
ground state to a polaronic ground state
does not contradict rigorous results 
on the absence of localization in this kind of model~\cite{ger91}.
Moreover, self-trapping does not imply
any change in the electronic density distribution.
If the electron is self-trapped by a local
lattice deformation, the resulting polaron 
is delocalized over the lattice and the polaron
appears only in correlations between electron and
lattice.

A measure of the polaronic character of the electron
is the correlation function
\begin{equation}
\chi_i\, = \, \frac{\langle n_i q_i \rangle}{2g\langle n_i \rangle} \, ,
\label{eq:qshift}
\end{equation}
where the index $i$ is either a site index $\ell$ on a chain
or $(x,y)$ on a square lattice.
Using~(\ref{eq:selfc}), one can also write
$\chi_i = \langle n_i q_i\rangle /\langle q_i\rangle$.
Therefore, it is clear that $|\chi_i| \leq 1$. 
In practice, we have found that $\chi_i$ takes only
positive value between 0 and 1.
For periodic boundary conditions, this function
is constant and differs from the function $\chi_{i,0}$ described
in Ref.~14 only by a factor of $L/2g$ ($L^2/2g$ in two dimensions).  
In open systems, the term $\langle n_i \rangle$ in the denominator is needed
to compensate for the inhomogeneous
density distribution. 
We have found that this function is almost constant,
except close to the lattice edges.
Here we report and discuss only values of $\chi_i$ obtained
in the central region of a lattice.

In Fig.~\ref{fig:crossover} we show our DMRG results
for $\chi_i$ as a function of the electron-phonon coupling $g$
for different values of $\omega/t$.
For small coupling $g$ our results tend
to the value predicted by the weak-coupling
perturbation theory.
For larger coupling, $\chi_i$ tends to 1 as predicted
by strong-coupling theory.
At intermediate coupling, one observes a rather sharp,
though continuous, transition from the weak-coupling 
to the strong-coupling value of $\chi_i$ as $g$ increases.
We think
that this transition marks the crossover from a quasi-free
electron ground state to a polaronic ground state.
The crossover roughly occurs when both conditions
$g>1$ and $g^2\omega \geq W/2$ are fulfilled,
in agreement with previous works~\cite{ran92,wel96,cap96}.
However, since the formation of polaron does not break
any symmetry and all ground state properties
are analytic functions of the parameters,
it is impossible to define critical values
$g_c$ and $\omega_c$ separating quasi-free electron
and polaronic regimes.
Unlike Capone et al.~\cite{cap96},
we have found that the crossover is always marked by a sharp increase
of $\chi_i$ in a small region of the plane ($g,\omega/t$),
even for large $\omega/t$.
The problem is that these authors have not normalized their
function $\chi_{i,0}$ by a factor $g$ as we do in Eq.~(\ref{eq:qshift}).
Therefore, they observe a quasi-linear dependence 
as a function of the electron-phonon coupling $g$,
which hides the sharp but small increase that we observe 
in Fig.~\ref{fig:crossover} at large phonon frequencies.

In two dimensions $\chi_i$ is smaller than in one dimension for the same
parameters $g$ and $\omega/t$. The crossover occurs at stronger coupling
because the band-width $W$ is larger in higher dimension and thus the 
condition $2g^2\omega > W$ is fulfilled for larger $g$.
However, differences between $\chi_i$ for one- and two-dimensional systems
diminish when the coupling increases (see results for $\omega=t$
in Fig.~\ref{fig:crossover}).

\subsection{Electronic kinetic energy}

One can obtain some insight about the electron state by calculating 
its kinetic energy (in units of the kinetic energy at $g=0$)
\begin{equation}
K \, = \frac{2t}{W} \sum_{\langle\ell,m\rangle} \left \langle c^{+}_{m}
c_{\ell} + c^{+}_{\ell} c_{m} \right \rangle  \, .
\label{eq:kinetic}
\end{equation}

Fig.~\ref{fig:kinetic1} and \ref{fig:kinetic2} show the evolution of the 
kinetic energy $K$ as a function of the electron-phonon coupling $g$
in the adiabatic and non-adiabatic regime, respectively. 
These results are qualitatively similar to recent exact diagonalization
results on small lattices~\cite{wel96}.
For weak coupling, $K$ is very close to 1. 
This means that the electron is barely affected by the
interaction with the phonons and remains essentially in the
same state as a free electron.
A further evidence for a quasi-free electron  ground state
is the good agreement between our DMRG results
and the second-order perturbation calculation in $g$,
at least as long as $g^2\omega < W/2$ or $g < 1$.
Therefore, we think that the electron is not trapped
by any lattice deformation in this regime but
simply drags a phonon cloud. 
We also note that for $\omega/t = t$ and $g=0.5$, 
static correlations $\langle n_i \, q_j\rangle$ decays over a few sites
(see Fig.~\ref{fig:correl1}(c))
while we find $K \approx 0.977$, which is not
compatible with an electron localized on a few sites.
This confirms that the spatial extent of lattice deformation
obtained from $\langle n_i \, q_j \rangle$ can be different from the 
localization length of the electron around
a lattice distortion. 

In the crossover region the kinetic energy decreases rapidly 
with increasing coupling. 
For large enough $g$ our DMRG
results tend to the values predicted by the second-order
strong-coupling theory.
The agreement between these results
is better for larger value of $\omega/t$
because the strong-coupling theory is a perturbative expansion
in $t/(g^2 \omega)$ and thus much more accurate
in the anti-adiabatic limit.
Also, our results confirm that the first-order strong-coupling
method, which predicts $K \sim \exp(-g^2)$, is a very poor
approximation for all values of $\omega/t$.
It is necessary to include at least
the second-order term in $t$ in the perturbative 
expansion to obtain reliable results.

In Fig.~\ref{fig:kinetic2} we can see that initially
$K$ decreases faster in one dimension than in two dimensions for
similar parameters.
Nevertheless, for large coupling $g$, our numerical results
and the strong-coupling theory show that $K$ converges
to the same values $\sim t/(g^2\omega)$ in both dimensions. 

Finally, we note that for $\omega=4t$ (see Fig.~\ref{fig:kinetic2}),
the combination of second-order weak- and strong-coupling
theory can reproduce our numerical results for all values of $g$
very accurately.
Therefore, these methods seems sufficiently accurate to study
the polaron problem in the anti-adiabatic limit and could be very 
useful in cases where numerical methods are not practical, for instance
in higher dimensions.
However, one should keep in mind that the strong-coupling theory
gives poor results in the crossover regime for smaller values of
$\omega/t$.

\section{Effective mass}
\label{sec:mass}

A polaron or a quasi-free electron with its phonon cloud
can be seen as an itinerant quasi-particle if its
effective band-width exceeds perturbations of its formation energy
by external forces.
Therefore, it is interesting to
compute parameters which describe its dynamics,
such as its effective mass $m^\star$.

The electronic density distribution 
shows that the electron or polaron
is delocalized over the lattice as a free particle.
We know that the band structure of a free particle 
in an open chain of length $L$
would be given by $E(k) = -2t^\star \cos(k)$ with
$k = z\pi/(L+1)$, where $z=1,2,3,..$ numbers the eigenstates.
The effective hopping term $t^\star$ is related to the effective mass
by $m^\star/m = t/t^\star$, where $m$ is the bare electron mass.
However, the polaron band structure is known to deviate
from this form because of the importance of effective
long-range hopping terms~\cite{wel96,ste96}.
Nevertheless, for large chains ($L >> 1$) we expect the
electronic excitation spectrum at low energy
to be given by
\begin{equation}
E(z,L) \, = E_\infty + t^\star \left ( \frac{z\pi}{L+1-2a} \right )^2 \, ,
\label{eq:energy}
\end{equation}
where $E_\infty$ is the ground state energy of an infinite chain and $a$ is
a parameter which account for the reduction of the effective system length
due to the repulsive effect the chain edge.
We can determine the parameters $E_\infty$, $t^\star$ and $a$ by calculating
different eigenenergies $E(z,L)$ with the DMRG and then fitting these results
to Eq.~(\ref{eq:energy}). 
In principle, we should vary $z$ in this equation and thus 
calculate the ground 
state and several excited states.
However, calculating accurate excitation energies with the DMRG 
is much more difficult than computing ground state energies. 
Moreover, the task of computing electronic excited states is complicated
by the intrusion of phononic excitations in the spectrum.
Therefore, we have obtained effective masses by fitting ground state
energies for several chain length $L$ to Eq.~(\ref{eq:energy}) with $z=1$.
This method only yields the effective mass at the bottom of the 
electronic or polaronic band but in this particular case gives results 
similar to those obtained by fitting excited state energies. 
We generally obtain excellent fit with this method as soon as
$L+1-2a > 10$.
We estimate that the error on our values for $m^\star$
is a few percents or smaller.
The value of $a$ which gives the best fit of the energy to
Eq.~(\ref{eq:energy})
is generally close to the value of $a$ which reproduces
the density distribution in Eq.~(\ref{eq:dens}).
Therefore, the behavior of the ground state energy as a function
of the system size confirms that the electron or polaron behaves like
a free particle on a chain of effective length $L-2a$ for all
values of the parameters $g$ and $\omega/t$.

In two dimensions we use a similar procedure.
The ground state energy for several square lattices of size $L\times L$
is fitted to Eq.~(\ref{eq:energy}) with $z=1$ and $2t^\star$ substituted
for $t^\star$.
The linear dimensions $L$ used in these calculations
were generally smaller than the chain lengths used in one-dimensional
systems.
Thus, the mass obtained for two-dimensional systems are less
accurate and we estimate that the relative error is $\leq 20\%$.

The structure (\ref{eq:energy}) of the electronic excitation spectrum
allow us to understand the main difficulty in applying
the DMRG method the polaron problem.
To determine the ground state accurately, we nee an absolute precision
which is better than the energy difference between the first excited
state ($z=2$) and the ground state ($z=1$).
Therefore, the relative error on the ground state energy must be smaller
than $\sim t^\star/(E_\infty L^2)$.
As the precision of our numerical method is limited by roundoff
errors, this condition imposes a constraint on the parameters
$g, \omega/t$ and $L$ for which we can find the ground state.
Using the strong-coupling theory results~\cite{hir83},
one can easily show that
for  $g \rightarrow \infty$,
$E_\infty \rightarrow -g^2\omega$ and $t^\star \rightarrow t \exp(-g^2)$.
Therefore, the minimal precision that we need goes
as $\sim t \exp(-g^2)/(L^2 g^2\omega)$ and becomes 
exceedingly small very rapidly with $g$.
In practice, 
we have been able to obtain the ground state
of chains with up to $L=16$ sites
for very heavy polaron ($t^\star/t \approx 10^{-4}$). 
Calculating the effective hopping accurately with (\ref{eq:energy}) 
requires a higher precision and thus is limited to
a smaller set of parameters. 
We can measure the effective hopping with a good accuracy
for $t^\star \geq 10^{-3} t$ using chains with up to
$L=30$ sites or square lattices with up to $10\times 10$ sites
at least.
In this paper we report results for the effective mass
(or hopping) in this range only.
Of course, in the quasi-free electron and crossover regimes,
where $t^\star \sim t$ and $E_\infty \leq 2t$ this problem
is less serious and we can study much larger systems.

In Fig.~\ref{fig:hopping1} we show the effective
hopping $t^\star$ calculated with our DMRG method
as well as the second order weak- and strong-coupling
results for $\omega/t=4$ in one dimension.
The good agreement between these results confirms both 
the accuracy of perturbative methods in the anti-adiabatic limit
and the validity of our method. 
We have found that our DMRG results also agree well with
the weak-coupling results in the quasi-free electron regime
for all values of $\omega/t$.
However, as $\omega/t$ decreases,
we observe differences between our results and the 
strong-coupling theory which becomes more and more important. 
The ratio between the values of $t^\star$ obtained with the DMRG
and the strong-coupling theory increases rapidly 
and can reach  $10^5$ for $\omega/t=0.1$. 
We think that this discrepancy is due to the limitation of the strong-coupling
theory which is a perturbative expansion in $t/(g^2\omega)$ and thus becomes
inaccurate for small $\omega/t$. 

In Fig.~\ref{fig:hopping2} we show the same results for $\omega/t=1$
together with the effective hopping obtained by 
a new Quantum Monte Carlo
(QMC) calculation~\cite{kor97}. 
There is qualitative agreement between DMRG and QMC results
but our values of $t^\star$ are always larger than those
obtained by QMC calculations. 
We can see that the DMRG method is more accurate than 
the QMC method at weak coupling.
We also note that QMC results are systematically lower than
the strong-coupling predictions.
On the other hand, we have
found that DMRG results are always larger than these
strong-coupling predictions. 
It is known that the second-order strong-coupling perturbation
theory underestimates the effective band-width for large 
coupling~\cite{ste96}.
Therefore, we think that this new QMC method underestimates
the effective hopping in the ground state of the Holstein model.

Finally, we show the effective mass $m^\star$ as a function of the 
electron-phonon coupling in Fig.~\ref{fig:mass}.
At finite coupling the quasi-free electron or polaron effective mass
is larger than the bare electron mass because of the phonon cloud
which must be dragged by the electron.
The sudden onset of self-trapping is marked by an abrupt increase of 
the effective mass.
However, the effective mass that we calculate is a ground state property
and its dependence on coupling constants
is smooth in agreement with exact theorems on the ground state
of the Holstein model~\cite{ger91}.
In the polaronic regime, the effective mass increases exponentially
with the coupling,
but in the adiabatic regime $\omega \leq t$
the mass enhancement is significantly smaller than 
the prediction of the first-order strong-coupling theory,
$m^\star/m=\exp(-g^2)$, 
as noted previously~\cite{wel96,ste96}.
The evolution of $m^\star$ is similar in one and two dimensions. 
The only difference is the shift of the crossover regime to a larger 
value of $g$ due to the variation of the bare electronic band-width $W$
as discussed in  the previous section.

As all ground state results are smooth at
the self-trapping transition 
we can not determine precisely
when a quasi-free electron becomes a polaron.
It is necessary to study excited states or dynamical properties
to find qualitative differences between both regimes~\cite{ran92,wel96}.
Nevertheless, our results show that for some parameters, for instance
$\omega=t$ and $g\approx 2-2.2$, the ground state is clearly a polaron
and the effective mass is relatively small, $m^\star/m \approx 10-100$.
Therefore, in the Holstein model there are polarons with an 
effective mass which is
much smaller than the prediction of the standard small polaron 
theory~\cite{hol59}.

\section{Conclusion}

In this paper we have presented a new DMRG approach to study
lattice systems including bosonic degrees of freedom.
The pseudo-site DMRG method is much more efficient than a 
standard approach using regular boson sites
and allow us to study large systems while
keeping up to 128 states in each bosonic Hilbert space.
We have successfully applied this method to the Holstein model
and we believe that it can be applied to any model including 
boson which can be studied with a standard DMRG method. 
A specific feature of the Holstein model is the absence
of direct interaction terms between bosons on different sites.
In models including such terms~\cite{pai96,car97}, one expects a decrease
of the pseudo-site DMRG performances because of the introduction
of additional long-range interactions between pseudo-sites.
Nevertheless, a pseudo-site approach is likely to be more
efficient than the standard approach even in this case.

The pseudo-site DMRG is just a method which efficiently handles 
the large number of states of a boson site.
A better approach would be to reduce
the number of states one needs to represent a boson site
using the key idea of DMRG.
In such an approach the reduced density matrix for a single
site is diagonalized to obtain a small set of optimized
states representing the boson site.
It has been shown that 3 optimized states per site give
results as accurate as with 10-100 states in exact diagonalizations
of the one-dimensional Holstein model at half-filling~\cite{zha97}.
Coupling this approach to the DMRG will further improve our capability
to perform numerical studies of systems including bosonic degrees
of freedom.

Using the pseudo-site DMRG method, we have studied the ground state 
of the one- and two-dimensional Holstein model with a single electron. 
We have been able to study all regimes of parameters
$g$ and $\omega/t$ in systems large enough to eliminate
finite size effects.
Our results are in good agreement with exact theorems,
perturbation theory predictions and the results of previous numerical works.
We have not found any qualitative differences between the 
one- and two-dimensional systems
after taking into account the doubling of the band-width in two dimensions 
compared to one dimension~\cite{ste96}.
In particular, in the weak coupling regime self-trapping does not
occur and the ground state is a quasi-free electron both in one and two 
dimensions~\cite{rae83}.
Several ground states properties
show a smooth but quite abrupt crossover from
a quasi-free electron to a polaronic ground state
as the electron-phonon coupling increases.
In particular, the crossover is signaled by a sharp 
increase of the effective mass, although
the mass enhancement can be much smaller than
predicted by the standard small polaron theory.

\acknowledgments
We wish to acknowledge the support of the Campus Laboratory
Collaborations Program of the University of California.
We acknowledge support from the NSF under Grant No. DMR-9509945,
and from the San Diego Supercomputer Center.
E.J. thanks the  Swiss National Science Foundation for 
financial support.


\begin{figure}
\caption{Ground state density distribution for $\omega=t$, $g=2.5$
in a 30-site chain.
The solid and dashed curves are given by equation (4.1) 
with $a$ = 1 and 0, respectively.}
\label{fig:density}
\end{figure}

\begin{figure}
\caption{Ground state density distribution for $\omega=t$, $g=2.2$
on a $9\times 9$ lattice.}
\label{fig:density2}
\end{figure}

\begin{figure}
\caption{Correlations $\chi_{10,j}$ 
between electron density and lattice
displacements on 20-site chains for various values
of $\omega/t$ and $g$.} 
\label{fig:correl1}
\end{figure}

\begin{figure}
\caption{Correlations $\chi(x,y)$ (see text)
between electron position and lattice deformation
on a $15\times 15$ lattice
with $\omega=0.2t$ and $g=0.1$. 
The electron position
is on the center of the lattice.}
\label{fig:correl2}
\end{figure}

\begin{figure}
\caption{Correlations $\chi(x,y)$ (see text)
between electron position and lattice deformation
on a $9\times 9$ lattice
with $\omega=t$ and $g=2.2$. 
The electron position
is on the center of the lattice.}
\label{fig:correl2b}
\end{figure}

\begin{figure}
\caption{Local electron-lattice correlation $\chi_i$ 
as a function of the electron-phonon coupling $g$
for $\omega / t  =$ 0.1 (circle), 0.2 (square), 1 (diamond)
and 4 (up triangle) in one dimension and for $\omega / t  =$ 1 
(down triangle) in two dimensions.
Open symbols are DMRG results.
Filled symbols show first-order weak-coupling perturbation results.}
\label{fig:crossover}
\end{figure}

\begin{figure}
\caption{Electronic kinetic energy 
as a function of the electron-phonon coupling $g$
in one-dimensional systems in the adiabatic regime.
Symbols are DMRG results.
Solid curves
show the second-order weak-coupling perturbation
results. Dashed curves are the predictions
of the second-order strong-coupling
expansion.}
\label{fig:kinetic1}
\end{figure}

\begin{figure}
\caption{Electronic kinetic energy 
as a function of the electron-phonon coupling $g$
in one-dimensional (1D) and two-dimensional (2D) systems
in the non-adiabatic regime.
Symbols are DMRG results.
Solid curves
show the second-order weak-coupling perturbation
results. Dashed curves are the predictions
of the second-order strong-coupling
expansion.}
\label{fig:kinetic2}
\end{figure}

\begin{figure}
\caption{Effective hopping integral $t^*$
as a function of the electron-phonon coupling $g$
for $\omega / t$~=~4 in one dimension. 
Symbols are DMRG results. The solid curve
is the second order weak-coupling perturbation
result. The dashed curve shows the second-order strong-coupling
expansion prediction.}
\label{fig:hopping1}
\end{figure}

\begin{figure}
\caption{Effective hopping integral $t^*$
as a function of the electron-phonon coupling $g$
for $\omega / t$~=~1 in one dimension. 
Diamonds are DMRG results.
The solid curve
is the second order weak-coupling perturbation
result. The dashed curve shows the second-order strong-coupling
expansion prediction.
Circles with error bars are QMC results.}
\label{fig:hopping2}
\end{figure}

\begin{figure}
\caption{Effective mass of the electron or polaron
as a function of the electron-phonon coupling $g$
for different values of $\omega / t$
in one-dimensional (1D) and two-dimensional (2D) systems.}
\label{fig:mass}
\end{figure}

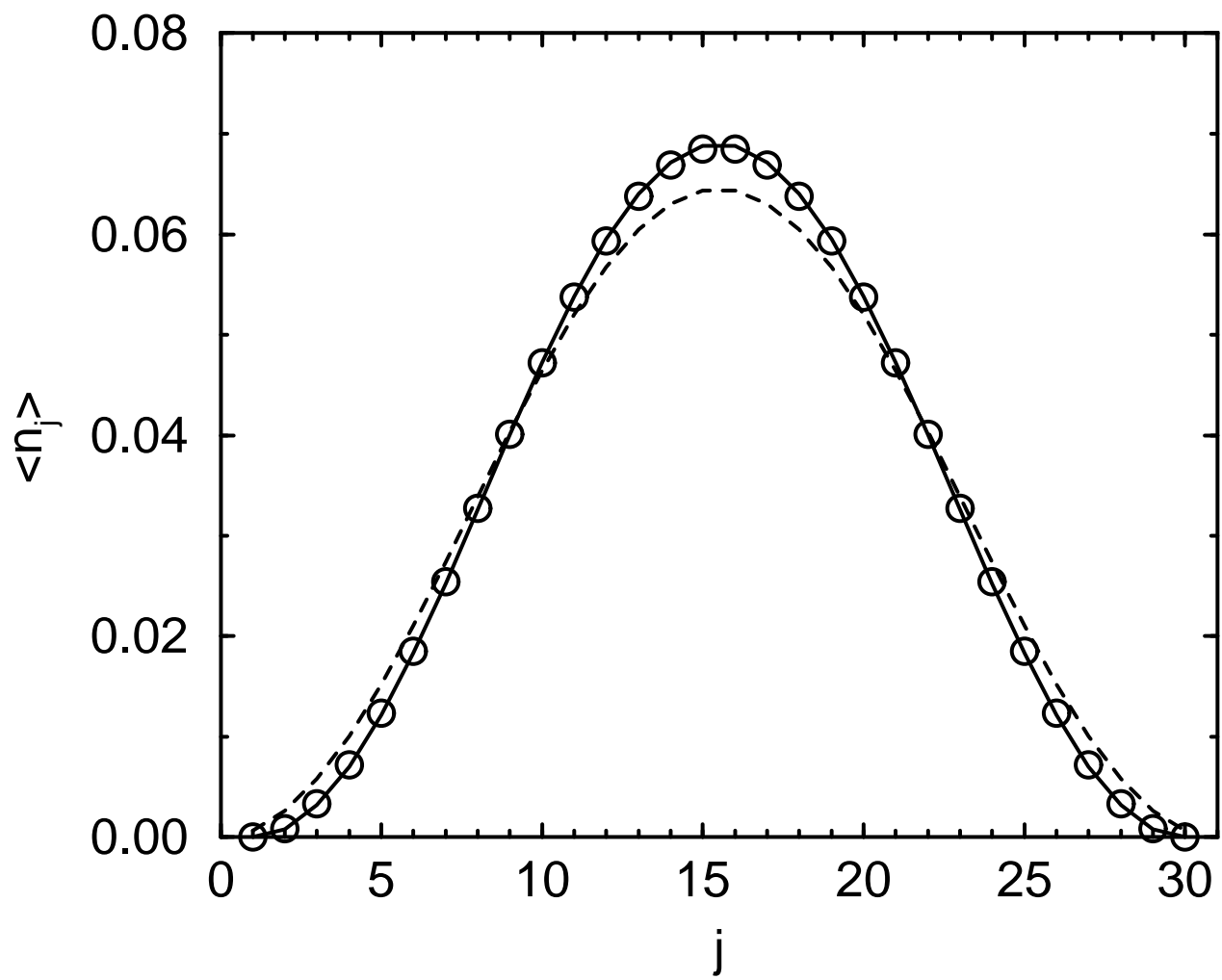
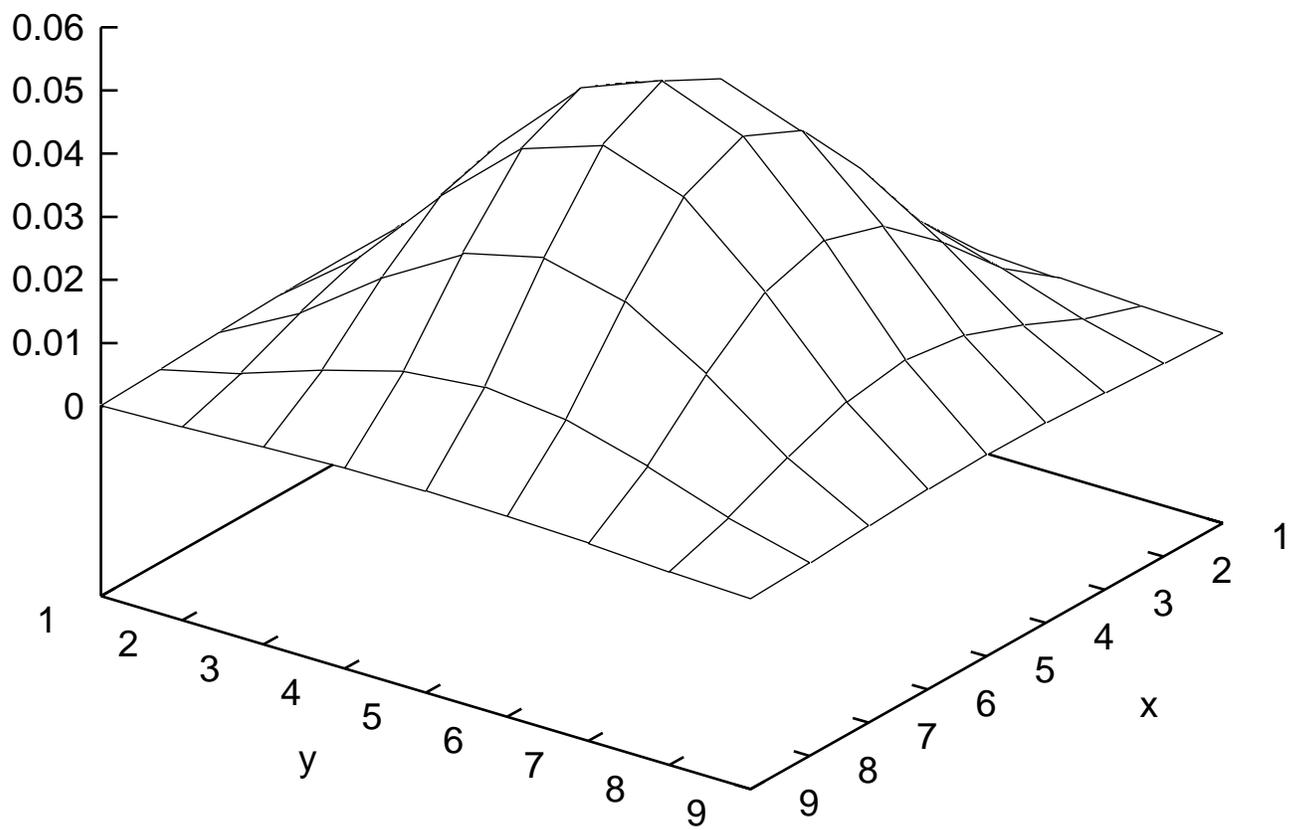
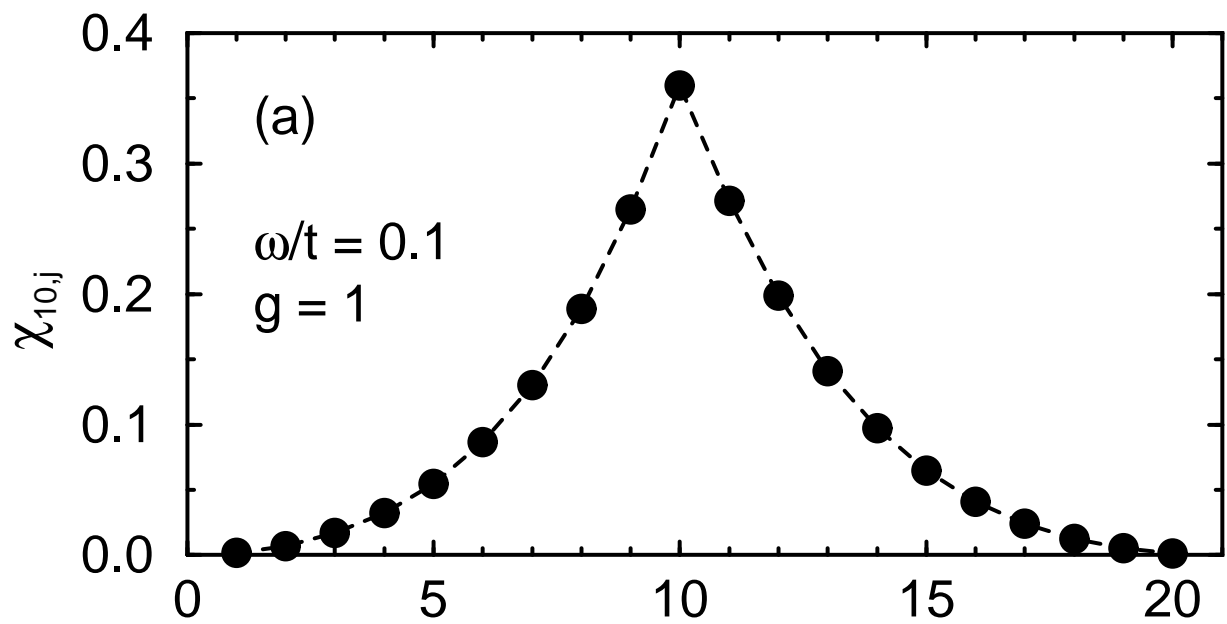
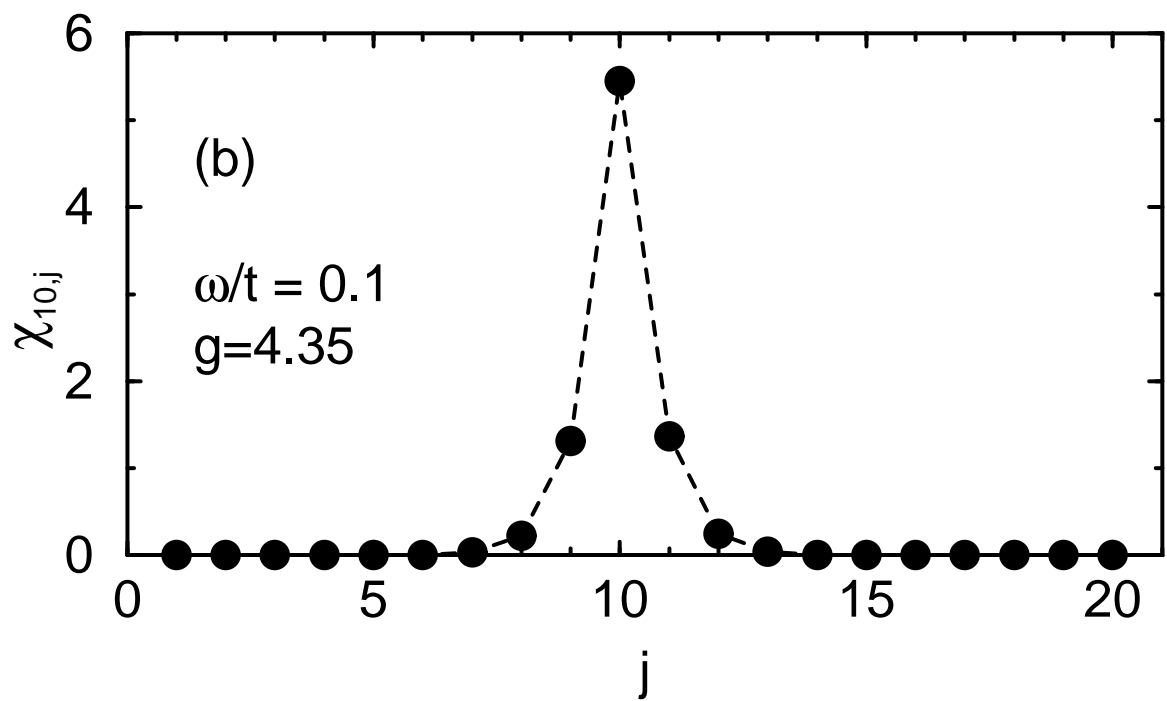
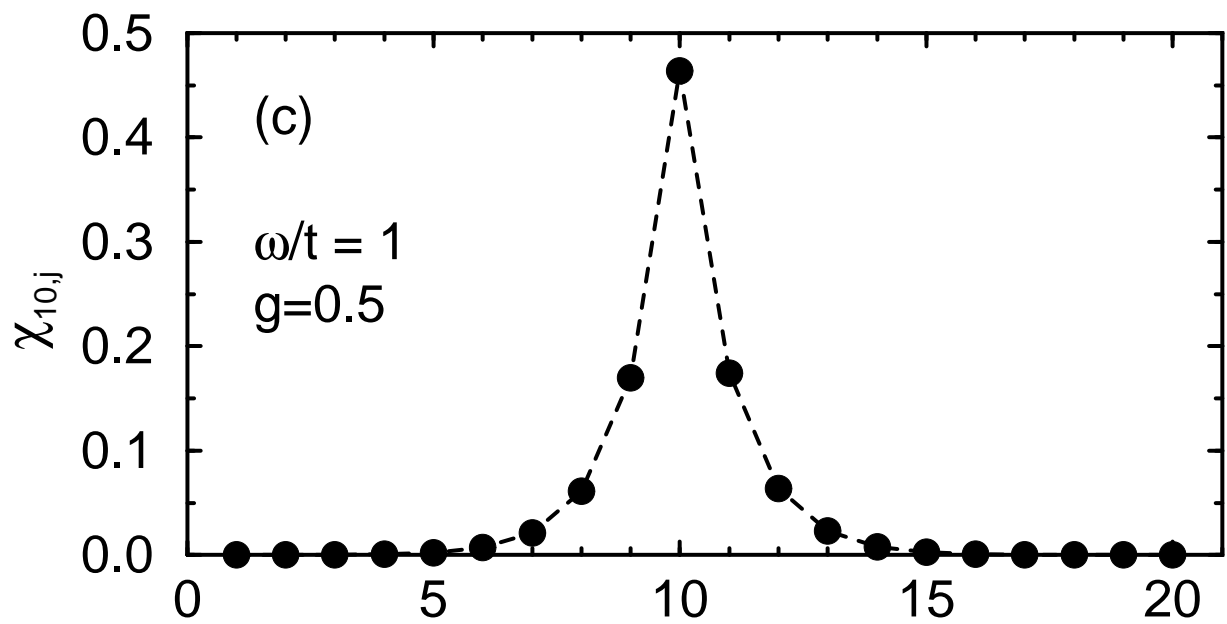
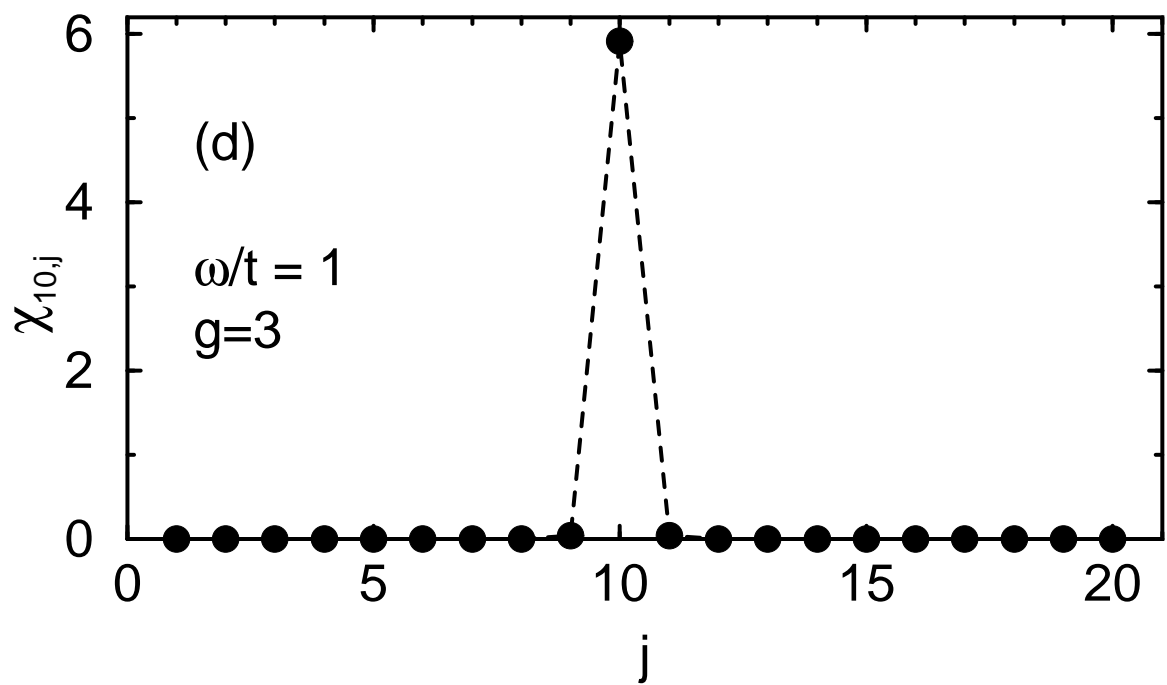
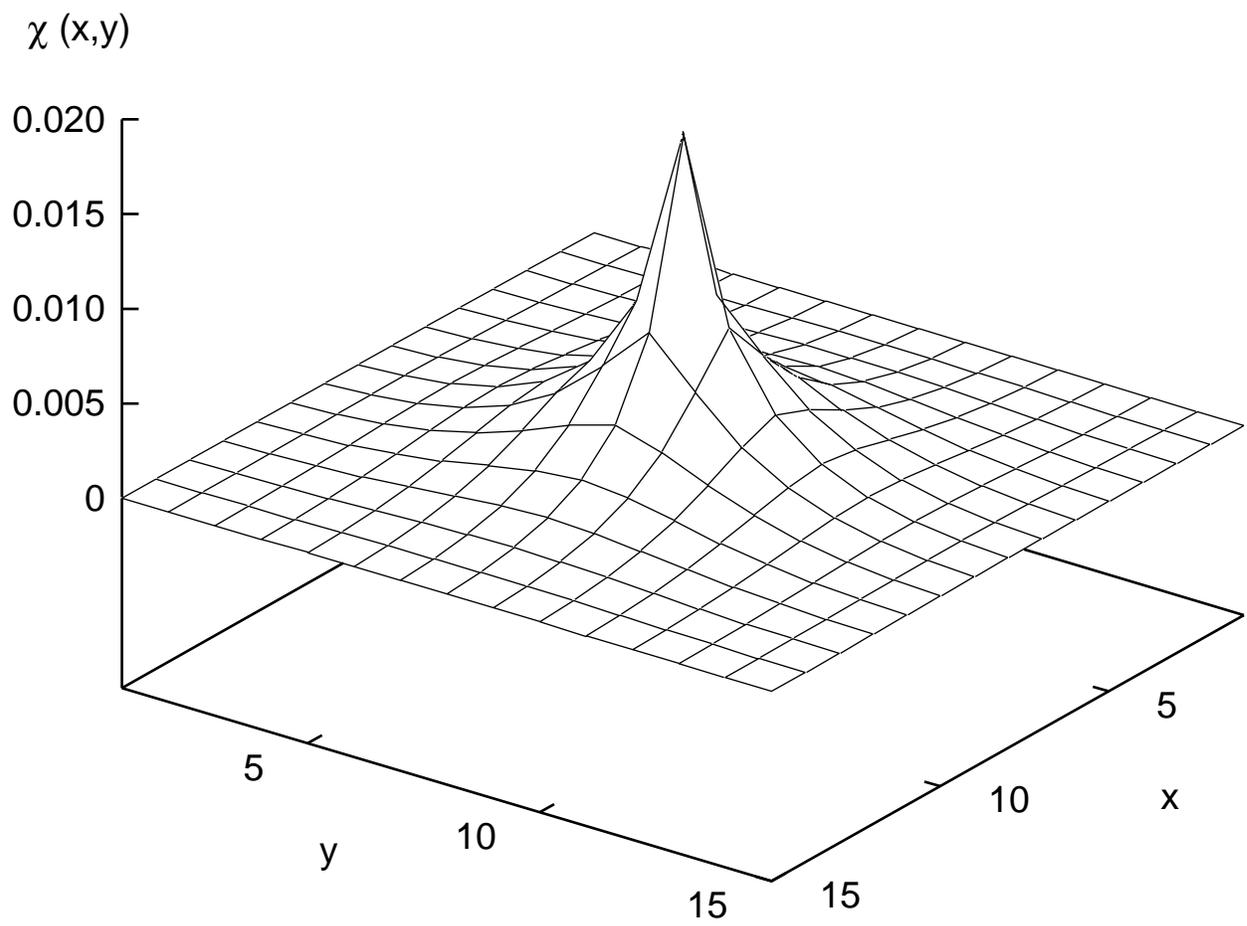
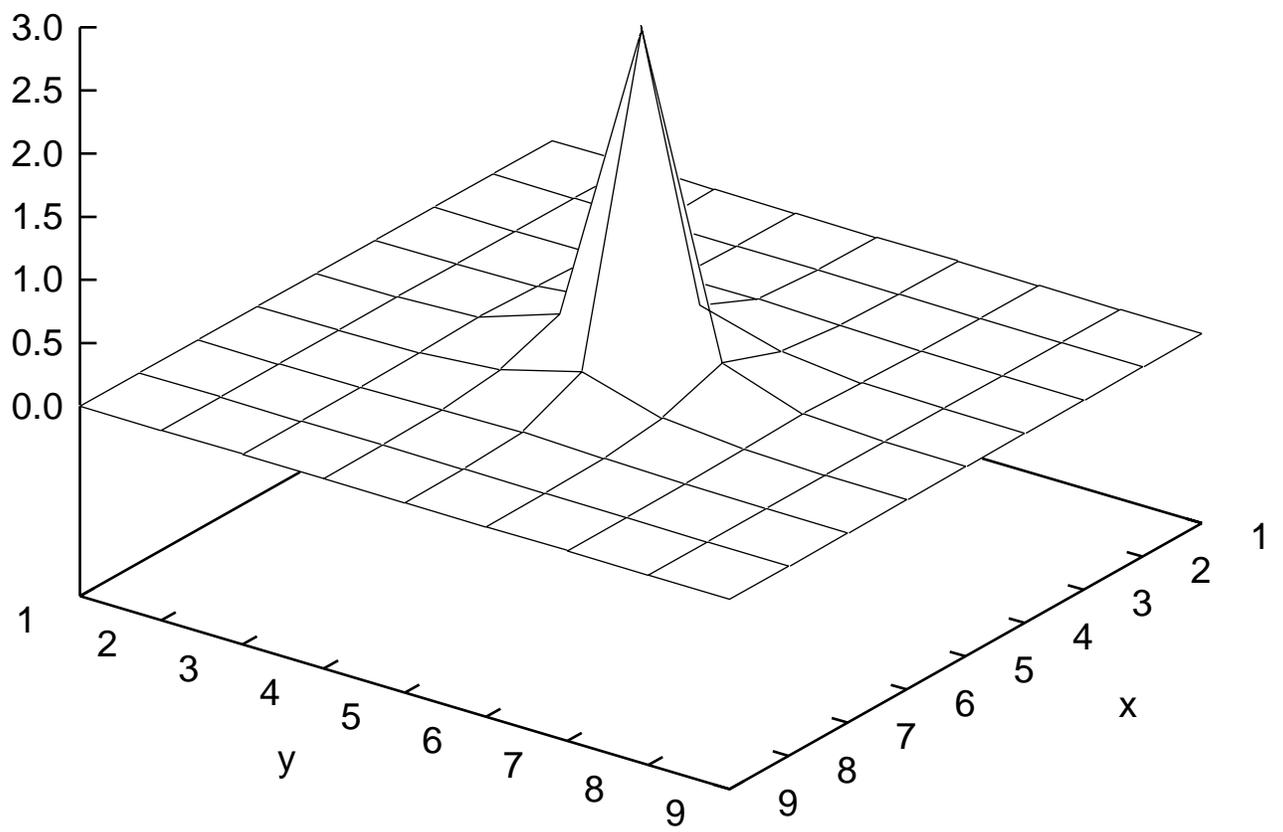
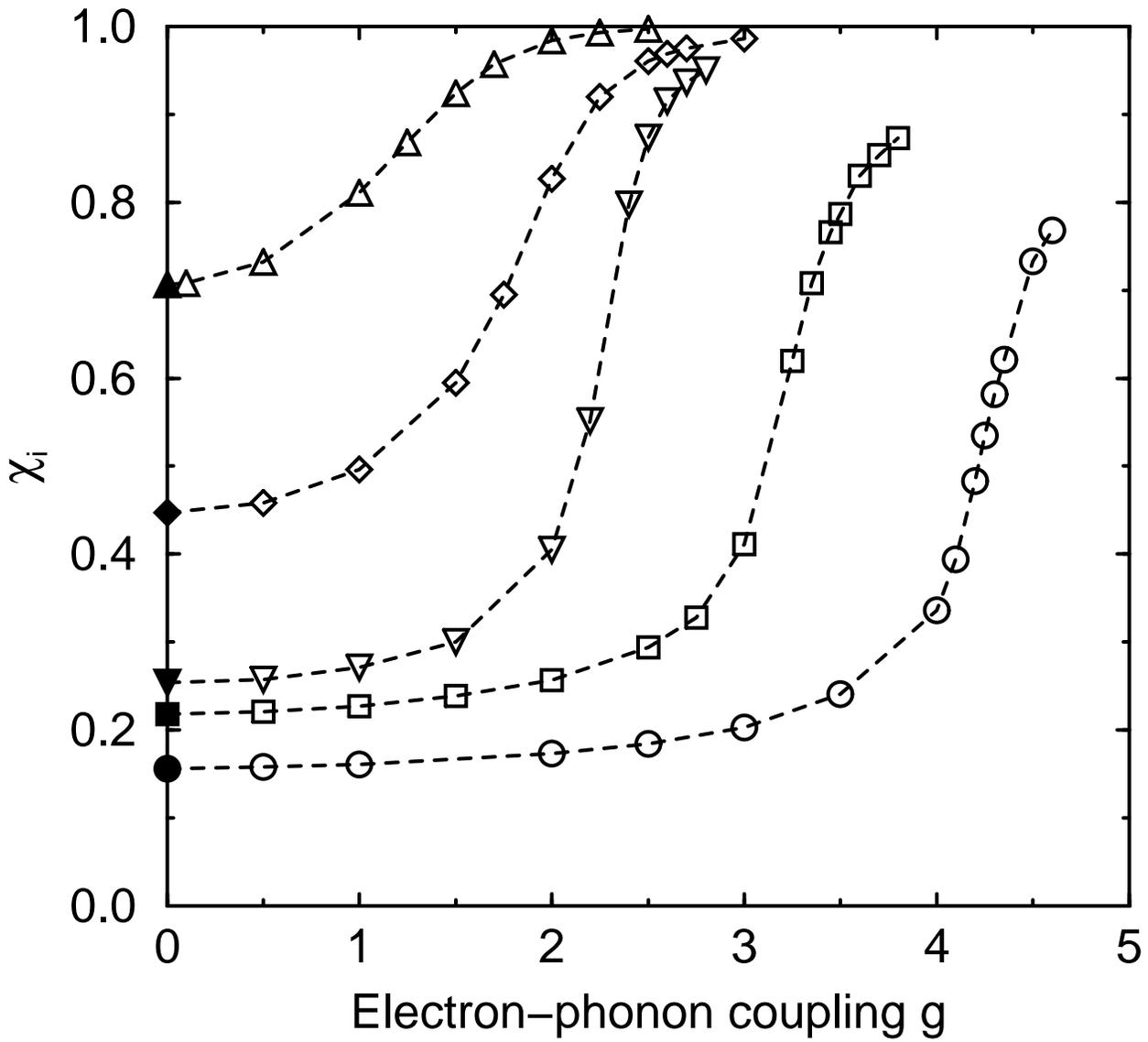
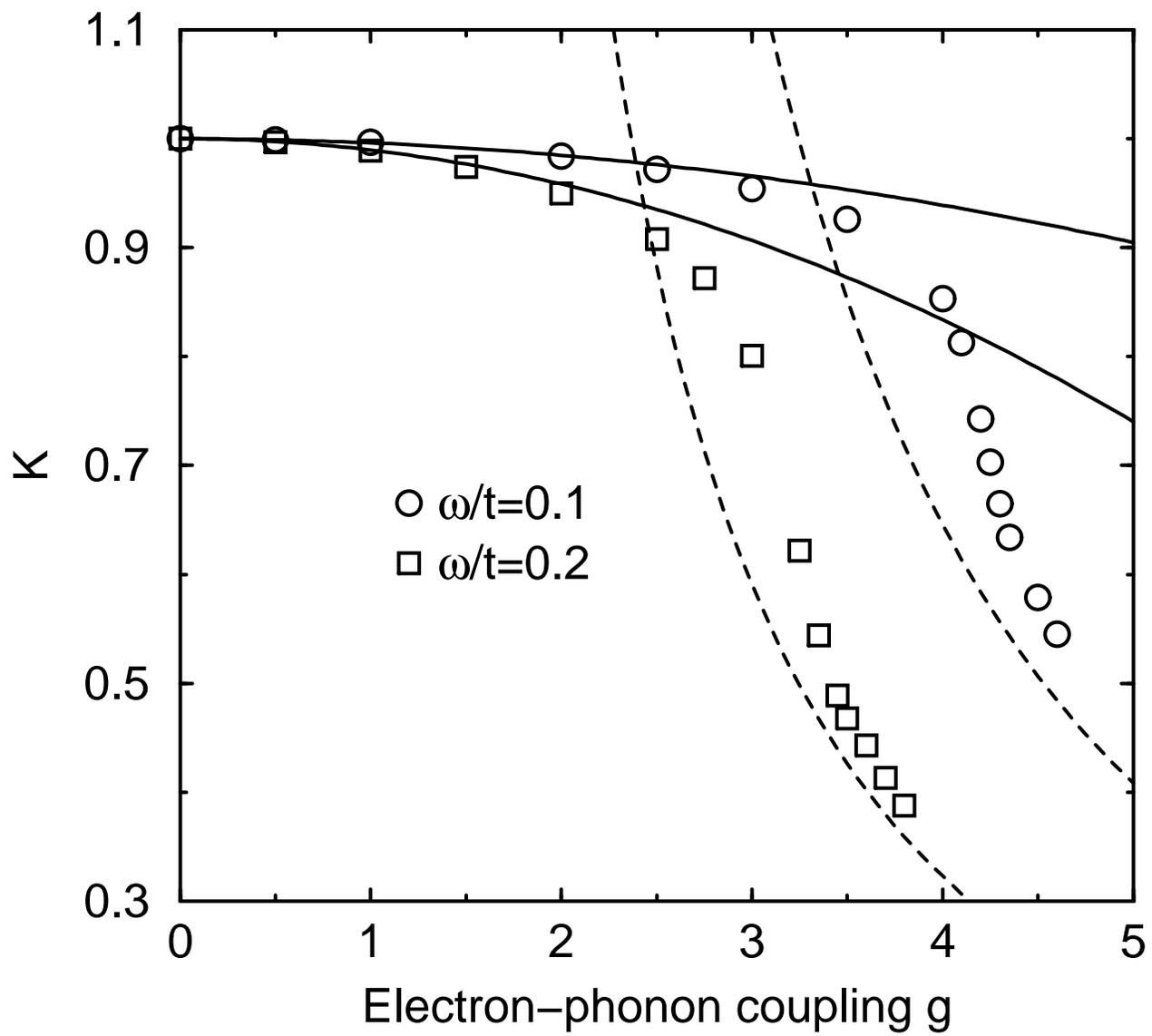
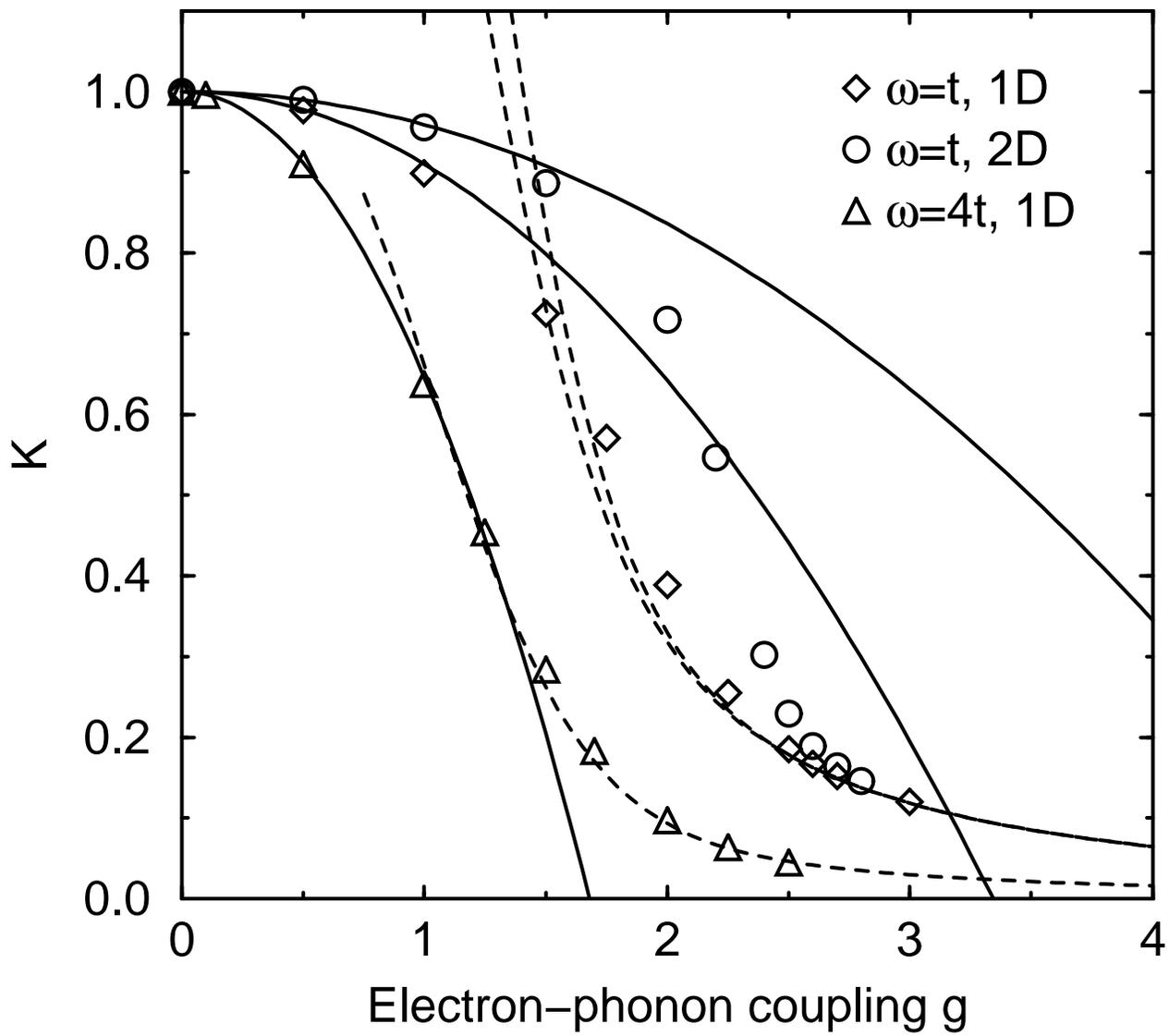
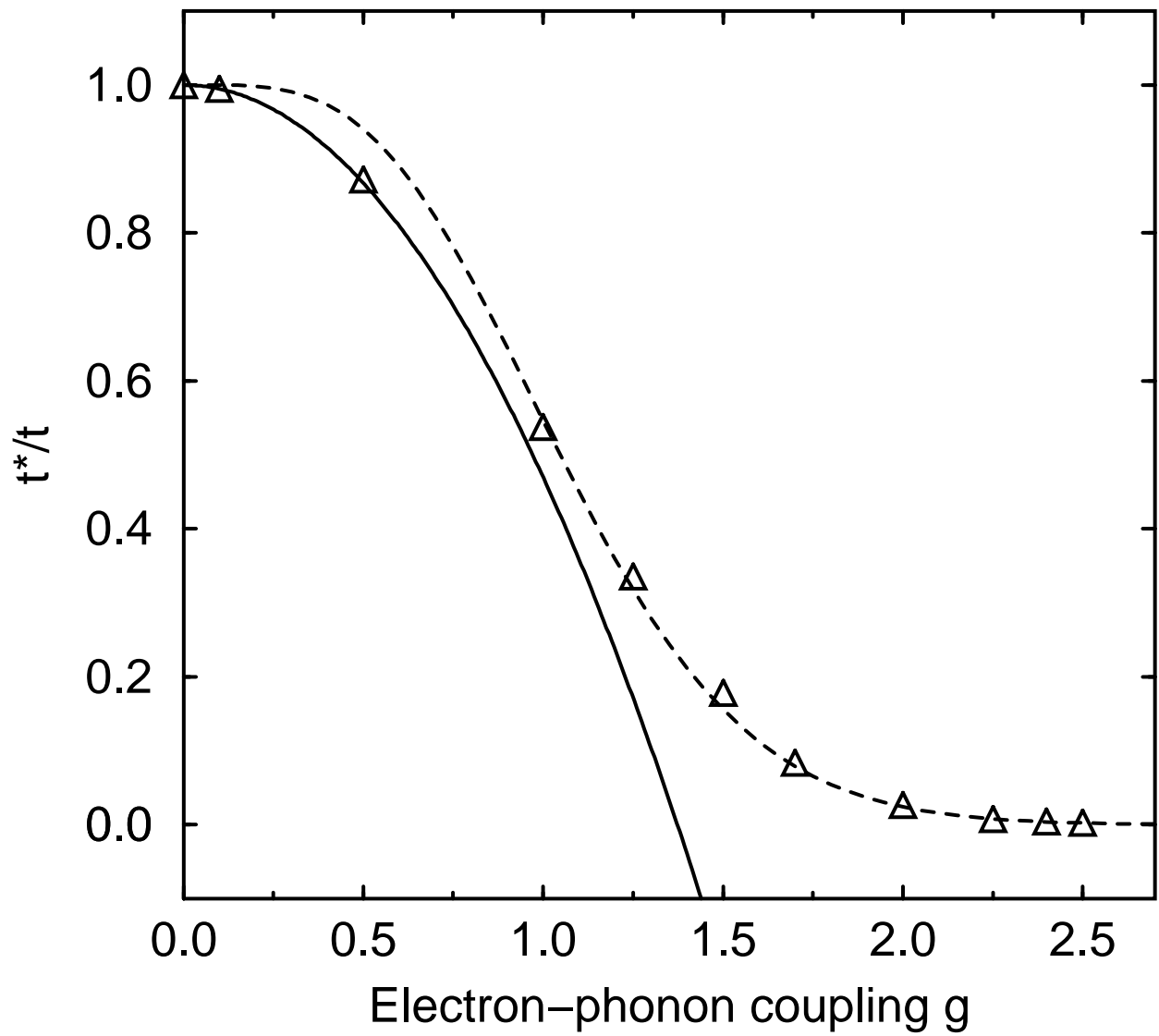
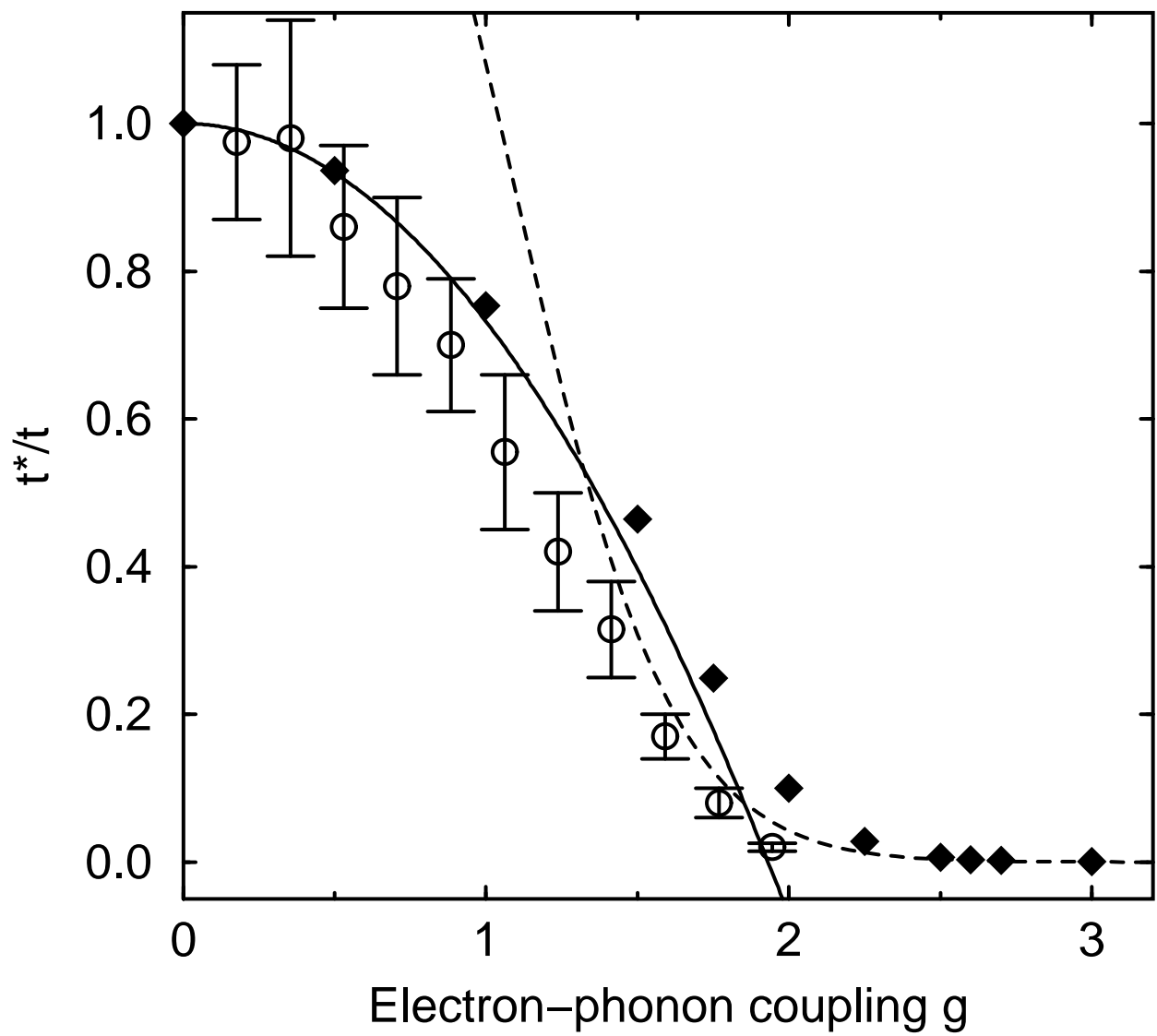
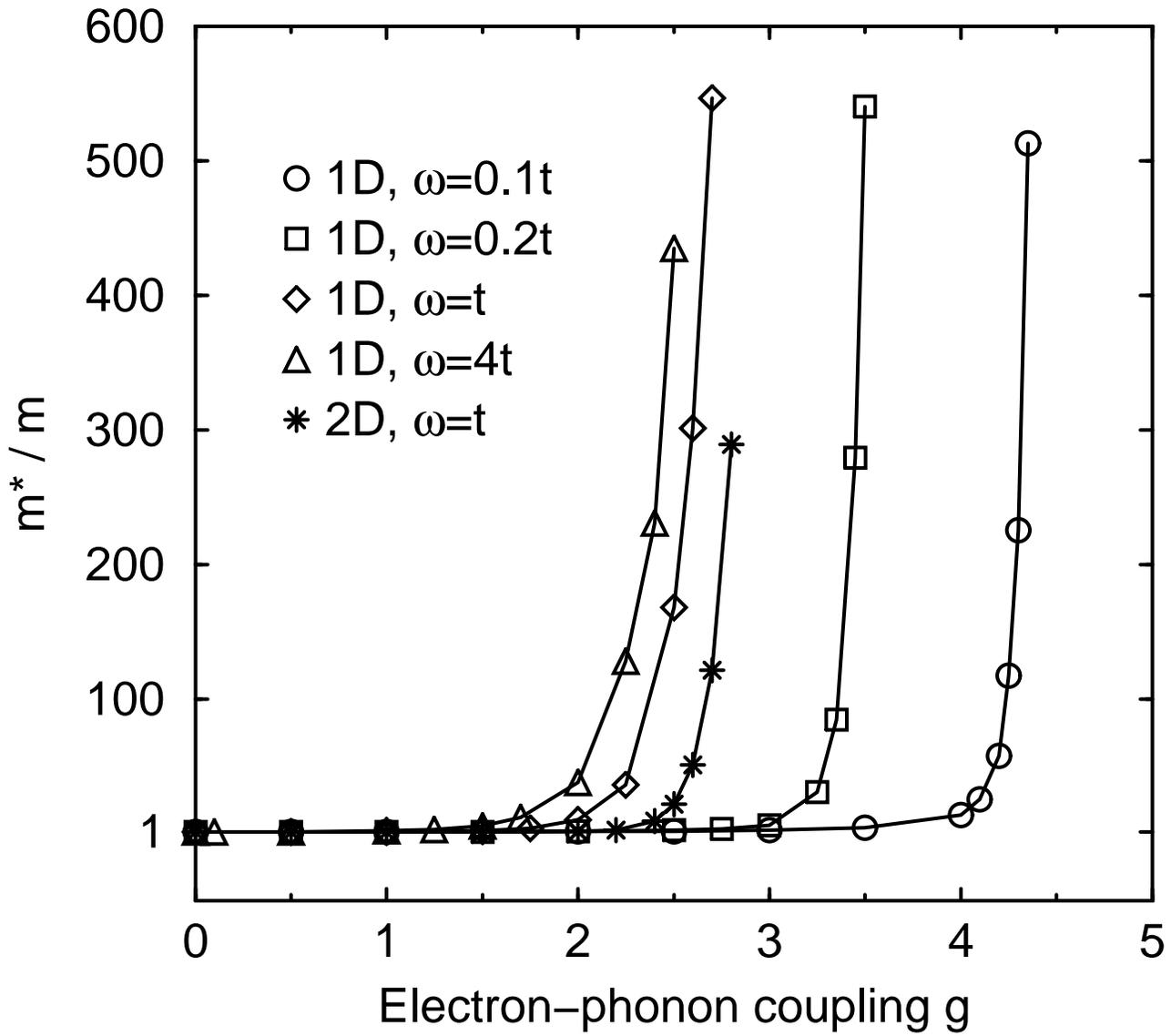
\end{document}